\documentclass[12pt,epsf]{article}
\usepackage{graphicx}
\usepackage{epsfig}
\begin{document}

\def\a{\alpha}
\def\b{\beta}
\def\c{\varepsilon}
\def\d{\delta}
\def\e{\epsilon}
\def\f{\phi}
\def\g{\gamma}
\def\h{\theta}
\def\k{\kappa}
\def\l{\lambda}
\def\m{\mu}
\def\n{\nu}
\def\p{\psi}
\def\q{\partial}
\def\r{\rho}
\def\s{\sigma}
\def\t{\tau}
\def\u{\upsilon}
\def\v{\varphi}
\def\w{\omega}
\def\x{\xi}
\def\y{\eta}
\def\z{\zeta}
\def\D{\Delta}
\def\G{\Gamma}
\def\H{\Theta}
\def\L{\Lambda}
\def\F{\Phi}
\def\P{\Psi}
\def\S{\Sigma}

\def\o{\over}
\def\beq{\begin{eqnarray}}
\def\eeq{\end{eqnarray}}
\newcommand{\gsim}{ \mathop{}_{\textstyle \sim}^{\textstyle >} }
\newcommand{\lsim}{ \mathop{}_{\textstyle \sim}^{\textstyle <} }
\newcommand{\vev}[1]{ \left\langle {#1} \right\rangle }
\newcommand{\bra}[1]{ \langle {#1} | }
\newcommand{\ket}[1]{ | {#1} \rangle }
\newcommand{\EV}{ {\rm eV} }
\newcommand{\KEV}{ {\rm keV} }
\newcommand{\MEV}{ {\rm MeV} }
\newcommand{\GEV}{ {\rm GeV} }
\newcommand{\TEV}{ {\rm TeV} }
\def\diag{\mathop{\rm diag}\nolimits}
\def\Spin{\mathop{\rm Spin}}
\def\SO{\mathop{\rm SO}}
\def\O{\mathop{\rm O}}
\def\SU{\mathop{\rm SU}}
\def\U{\mathop{\rm U}}
\def\Sp{\mathop{\rm Sp}}
\def\SL{\mathop{\rm SL}}
\def\tr{\mathop{\rm tr}}

\def\IJMP{Int.~J.~Mod.~Phys. }
\def\MPL{Mod.~Phys.~Lett. }
\def\NP{Nucl.~Phys. }
\def\PL{Phys.~Lett. }
\def\PR{Phys.~Rev. }
\def\PRL{Phys.~Rev.~Lett. }
\def\PTP{Prog.~Theor.~Phys. }
\def\ZP{Z.~Phys. }

\newcommand{\bear}{\begin{array}}  
\newcommand {\eear}{\end{array}}
\newcommand{\la}{\left\langle}  
\newcommand{\ra}{\right\rangle}
\newcommand{\non}{\nonumber}  
\newcommand{\ds}{\displaystyle}
\newcommand{\red}{\textcolor{red}}
\def\ubl{U(1)$_{\rm B-L}$}
\def\REF#1{(\ref{#1})}
\def\lrf#1#2{ \left(\frac{#1}{#2}\right)}
\def\lrfp#1#2#3{ \left(\frac{#1}{#2} \right)^{#3}}
\def\OG#1{ {\cal O}(#1){\rm\,GeV}}


\baselineskip 0.7cm

\begin{titlepage}

\begin{flushright}
UT-09-15\\
IPMU 09-0065
\end{flushright}

\vskip 1.35cm
\begin{center}
{\large \bf
Cosmic-ray Electron and Positron Excesses from Hidden Gaugino Dark Matter
}
\vskip 1.2cm
Satoshi Shirai$^{1,2}$, Fuminobu Takahashi$^2$ and  T. T. Yanagida$^{2,1}$
\vskip 0.4cm

{\it $^1$  Department of Physics, University of Tokyo,\\
     Tokyo 113-0033, Japan\\
$^2$ Institute for the Physics and Mathematics of the Universe, 
University of Tokyo,\\ Chiba 277-8568, Japan}

\vskip 1.5cm

\abstract{ We study a scenario that a hidden gaugino dark matter
  decays into the standard-model particles (and their supersymmetric partners) through a kinetic mixing
  with the gaugino of a U(1)$_{\rm B-L}$ broken at a scale close to
  the grand unification scale.  We show that decay of the hidden
  gaugino can explain excesses in the cosmic-ray electrons and
  positrons observed by PAMELA and Fermi.  }
\end{center}
\end{titlepage}

\setcounter{page}{2}

\section{Introduction}
\label{sec:1}

The cosmic-ray electrons and positrons have attracted much attention
since the PAMELA collaboration~\cite{Adriani:2008zr} released the data
showing rapid growth in the positron fraction from several tens GeV up
to about $100$\,GeV.  Recently, the cosmic-ray electron plus positron
flux was measured with the Fermi satellite~\cite{Collaboration:2009zk}
with significantly improved statistics. The Fermi data shows that the
$(e^-+e^+)$ spectrum falls as $E^{-3.0}$ over energies between
$20$\,GeV and $1$\,TeV without prominent spectral features.  The
H.E.S.S. collaboration also measured the cosmic-ray $(e^-+e^+)$
spectrum from $340$\,GeV up to several
TeV~\cite{Collaboration:2008aaa}, suggesting that the spectrum
steepens above $1$\,TeV.  The Fermi and H.E.S.S. results are in
agreement with each other where the energy of the two data overlaps.
Combining the PAMELA, Fermi and H.E.S.S. results, therefore, it is
likely that there is an excess in the electron and positron flux above
several tens GeV up to $1$\,TeV.

We have recently presented a scenario that thermal relic Wino dark
matter (DM) of mass about $3$\,TeV, decaying through an R-parity
violating operator ${\bar e}LL$, naturally accounts for the PAMELA and
Fermi excesses simultaneously~\cite{Shirai:2009fq}.  In the model, the
magnitude of the R-parity breaking as well as the Wino mass are
closely tied to the gravitino mass, $m_{3/2}$, of ${\cal
  O}(10^3)$\,TeV. Interestingly enough, the lifetime of the Wino DM
naturally becomes of ${\cal O}(10^{26})$\,seconds, which is suggested
by observation to account for the electron/positron excess.  The only
drawback of this scenario might be that all supersymmetric (SUSY)
particles must have masses heavier than several TeV and therefore
beyond the reach of LHC.

In this paper, we consider a hidden gaugino of an unbroken U(1) gauge
symmetry as a candidate for
DM~\cite{Shirai:2009kh,Ibarra:2008kn}. Since the longevity of DM
originates from its extremely weak interactions with the standard
model (SM) particles, the SUSY particles in the SM sector
can have masses well below the DM mass, possibly within the reach of
LHC.  As pointed out in Ref.~\cite{Shirai:2009kh}, the decay of the
hidden gaugino proceeds through a kinetic mixing with a U(1)$_{\rm
  B-L}$ gaugino living in the bulk (see also Ref.~\cite{Chen:2008yi}).
The decay rate suppressed by the U(1)$_{\rm B-L}$ breaking scale
provides the desired magnitude of the lifetime. In addition, the
hidden gaugino DM will mainly decay into a lepton and slepton pair, if
the squarks are substantially heavier than the sleptons. Therefore the
decay process can be lepto-philic in concordance with the absence of
the excess in the antiproton fraction~\cite{Adriani:2008zq}.  In
Ref.~\cite{Shirai:2009kh}, we considered a case that the hidden
gaugino decays universally into a lepton and slepton pair in the three
generations. In this paper, we study more generic decay processes such
as that into the third generation as well as three-body decays with a
virtual slepton exchange. We will show the decays of the hidden
gauginos can explain the anomalous excesses observed by PAMELA and
Fermi.

This paper is organized as follows. In Sec.~\ref{sec:2} we briefly
describe our hidden gaugino DM model.  The predicted positron fraction
and the electron spectrum will be shown in Sec.~\ref{sec:3}. The last
section is devoted to discussion and conclusions.

\section{Model}
\label{sec:2}
In this section we will briefly describe the model proposed in
Ref.~\cite{Shirai:2009kh}.  The reader is referred to the original
reference for more details.

Suppose that a hidden U(1) gauge multiplet $(\lambda_{H},A_{H},
D_{H})$ is confined on a brane, which is geometrically separated from
the brane on which the SUSY SM (SSM) particles reside, in a set up
with an extra dimension. We introduce a U(1)$_{\rm B-L}$ gauge
multiplet in the bulk so that those two sectors are in contact only
through a kinetic mixing of the U(1)$_{\rm B-L}$ and hidden U(1)
multiplets. The mixing is written as
\beq
{\cal L }_K &=& \frac{1}{4}\int d^2\theta (W_{ H}W_{H} + W_{\rm B-L}W_{\rm B-L} + 2 \kappa W_{H}W_{\rm B-L})  
+ {\rm h.c.}, \non\\
&\supset&-i \left( \bar \lambda_{H} \bar \sigma^\mu \partial_\mu \lambda_{ H}
+ \bar \lambda_{\rm B-L} \bar \sigma^\mu \partial_\mu \lambda_{\rm B-L}
+  \kappa \bar \lambda_{H} \bar \sigma^\mu \partial_\mu \lambda_{\rm B-L}
+  \kappa \bar \lambda_{\rm B-L} \bar \sigma^\mu \partial_\mu \lambda_{H}
\right),
\label{eq:mixing}
\eeq
where $\kappa$ is a kinetic mixing parameter of ${\cal O}(0.1)$.  Using the
same notation $\lambda_H$ for the hidden gaugino in the mass
eigenstate, its interaction with the SSM particles can be expressed
as~\cite{Shirai:2009kh}
\beq
{\cal L}_{\rm int}\;\simeq \;
-\sqrt{2}g_{\rm B-L} Y_{\psi} \kappa \lrfp{m}{M}{2} \lambda_H\phi^{*}_{\rm SSM} \psi_{\rm SSM}  + {\rm h.c.}, 
 \label{eq:int}
 \eeq%
 where $g_{\rm B-L}$ denotes the U(1)$_{\rm B-L}$ gauge coupling,
 $Y_{\psi}$ is the (B$-$L) number of $\phi$ and $\psi$, $m$ represents
 a soft SUSY breaking Majorana mass of $\lambda_H$, and $M (\equiv 4
 g_{\rm B-L} v_{\rm B-L})$ is the mass of $\lambda_{\rm B-L}$ arising
 from the spontaneous breaking of U(1)$_{\rm B-L}$ at a scale $v_{\rm
   B-L}$.  If the masses of $\phi$ and $\psi$ are much smaller than
 $m$, the lifetime of $\lambda_H$ is estimated to be
 \beq \Gamma_{\rm DM}^{-1} ( \lambda_H \to \psi+\phi)\sim 10^{24}~{\rm
   sec}~ g^{-2}_{\rm B-L} Y^{-2}_{\psi}\kappa^{-2}
 \left(\frac{m}{3~{\rm TeV}} \right)^{-5} \left(\frac{M}{10^{16}
     ~\GEV}\right)^{4}\frac{1}{C_{\psi}}, \eeq
 where $C_{\psi}$ is a color factor of $\psi$, i.e., $3$ for quarks
 and $1$ for leptons.  On the other hand, if the mass of $\phi$ is
 larger than $m$, the decay proceeds with a virtual exchange of
 $\phi$, leading to
\beq
\Gamma_{\rm DM}^{-1} ( \lambda_H \to \psi + {\psi}^{\prime *} + \tilde{\chi} )\sim
10^{26}~{\rm sec}~  g^{-2}_{\rm B-L} Y^{-2}_{\psi}\kappa^{-2}  \lrfp{m_\phi}{m}{4}
\left(\frac{m}{3~{\rm TeV}} \right)^{-5} \left(\frac{M}{10^{16} ~\GEV}\right)^{4}\frac{1}{C_{\psi}},
\eeq
where we have assumed that the main decay of $\phi$ is into $
{\psi}^{\prime *}$ and a SM gaugino $ \tilde{\chi}$.

It is quite remarkable that the hierarchy between the B$-$L  breaking
scale $\sim 10^{16}$ GeV~\footnote{ The seesaw mechanism~\cite{seesaw}
  for neutrino mass generation suggests the Majorana mass of the
  (heaviest) right-handed neutrino at about the GUT scale. Such a
  large Majorana mass can be naturally provided if the U(1)$_{\rm
    B-L}$ symmetry is spontaneously broken at a scale around $10^{16}$
  GeV.  } and the SUSY breaking mass of the hidden gaugino of
${\cal O}(1)$\,TeV naturally leads to the lifetime of ${\cal O}(10^{26})$ seconds
that is needed to account for the electron/positron excess.  Note also
that the longevity of the $\lambda_H$ DM arises from the geometrical
separation and the hierarchy between $M$ and $m$, not from
conservation of some discrete symmetry such as the R parity.

Throughout this paper we assume that the R parity is preserved, and
that the lightest supersymmetric particle (LSP) is the lightest neutralino in the SSM which also
contributes to the DM density. For the moment we assume that
$\lambda_H$ is the dominant component of DM, while the abundance of
the neutralino LSP is negligible. Even if this is not the case, the
prediction on the cosmic-ray fluxes given in the next section can
remain unchanged, since the fraction of $\lambda_H$ in the total DM
density can be traded off with the lifetime, as long as the fraction
is larger than about $10^{-10}$.  We will come back to this issue in
Sec.~\ref{sec:4}.

\section{Electron and positron excesses from the decaying hidden-gaugino DM}
\label{sec:3}
As we have seen in the last section, $\lambda_H$ has a very long
lifetime and decays into the SSM particles through a small mixing with
the $\lambda_{\rm B-L}$.  The decay of $\lambda_H$ causes the SUSY
cascade decays, emitting high energy SM particles.  If the squarks are
heavier than $\lambda_H$, the hadronic decay can be suppressed, which
results in a small amount of antiprotons and
photons~\cite{Arvanitaki:2008hq}.  We assume that it is the case.

We fix in the present analysis the masses of the $\lambda_H$ and the
neutralino LSP to be $3$ TeV and $200$ GeV, respectively.  In general,
SUSY cascade decays are very complicated. To simplify the analysis, we
focus on the following three extreme cases.
\begin{description}
\item[Case I] [{\bf Universal decay into  a lepton and slepton pair}]:
The $\lambda_H$  decays into the three lepton and slepton pairs at the same rate, and the sleptons subsequently 
decay into LSP + charged lepton.\footnote{
If the LSP is the Wino, the slepton may also decay into neutrino and charged Wino with a large branching fraction.
} We set that the lifetime of  $\lambda_H$  is $9\times 10^{25}$ sec., and
$m_{\tilde{e}_R}=m_{\tilde{\mu}_R}=m_{\tilde{\tau}_1}=2.5$ TeV and the other sfermion particles are heavier than $\lambda_H$.
\item[Case II] [{\bf Decay into a tau and stau pair}]: The $\lambda_H$
  decays into a tau and stau pair, and the stau decays into LSP + tau.
  The lifetime of $\lambda_H$ is $6\times 10^{25}$ sec., and
  $m_{\tilde{\tau}_1}=2.5$ TeV and the other sfermion particles are
  heavier than $\lambda_H$.
\item[Case III] [{\bf Three-body decay into the lepton, anti-lepton
    and neutralino}]: The $\lambda_H$ decays into the
  $e^{+}+e^{-}+{\rm LSP}$, $\mu^{+}+\mu^{-}+{\rm LSP}$,
  $\tau^{+}+\tau^{-}+{\rm LSP}$ at the same rate. The lifetime of
  $\lambda_H$ is $1.1\times 10^{26}$ sec.  All sfermion particles are
  heavier than $\lambda_H$.  As for the matrix element of the DM
  decay, we approximate that the sleptons are much heavier than
  $\lambda_h$, i.e., $m_{\tilde{\ell}}\gg m$.
\end{description}

\begin{figure}[ht]
\begin{center}
\epsfig{file=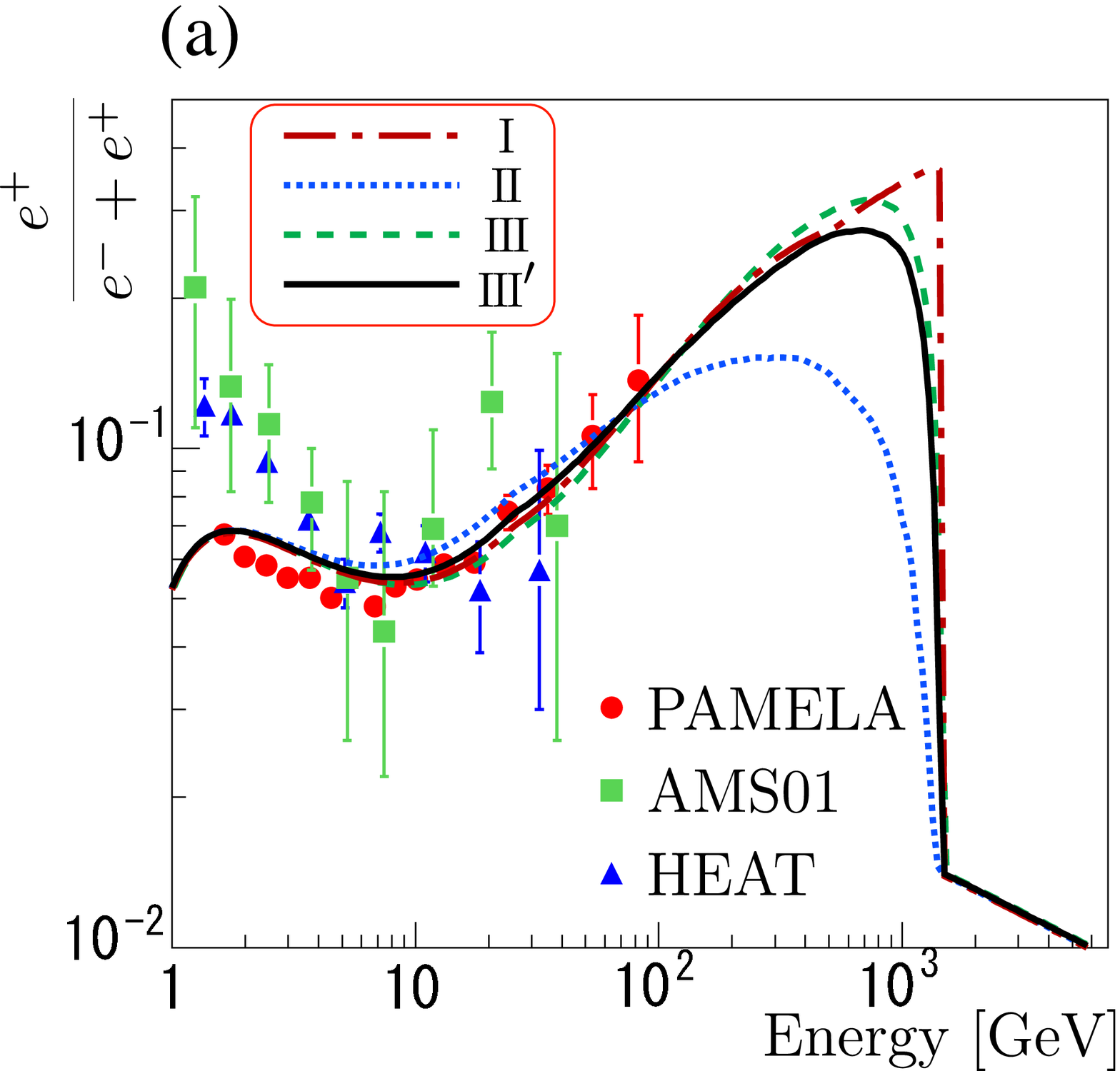 ,scale=.50,clip}\\
\epsfig{file=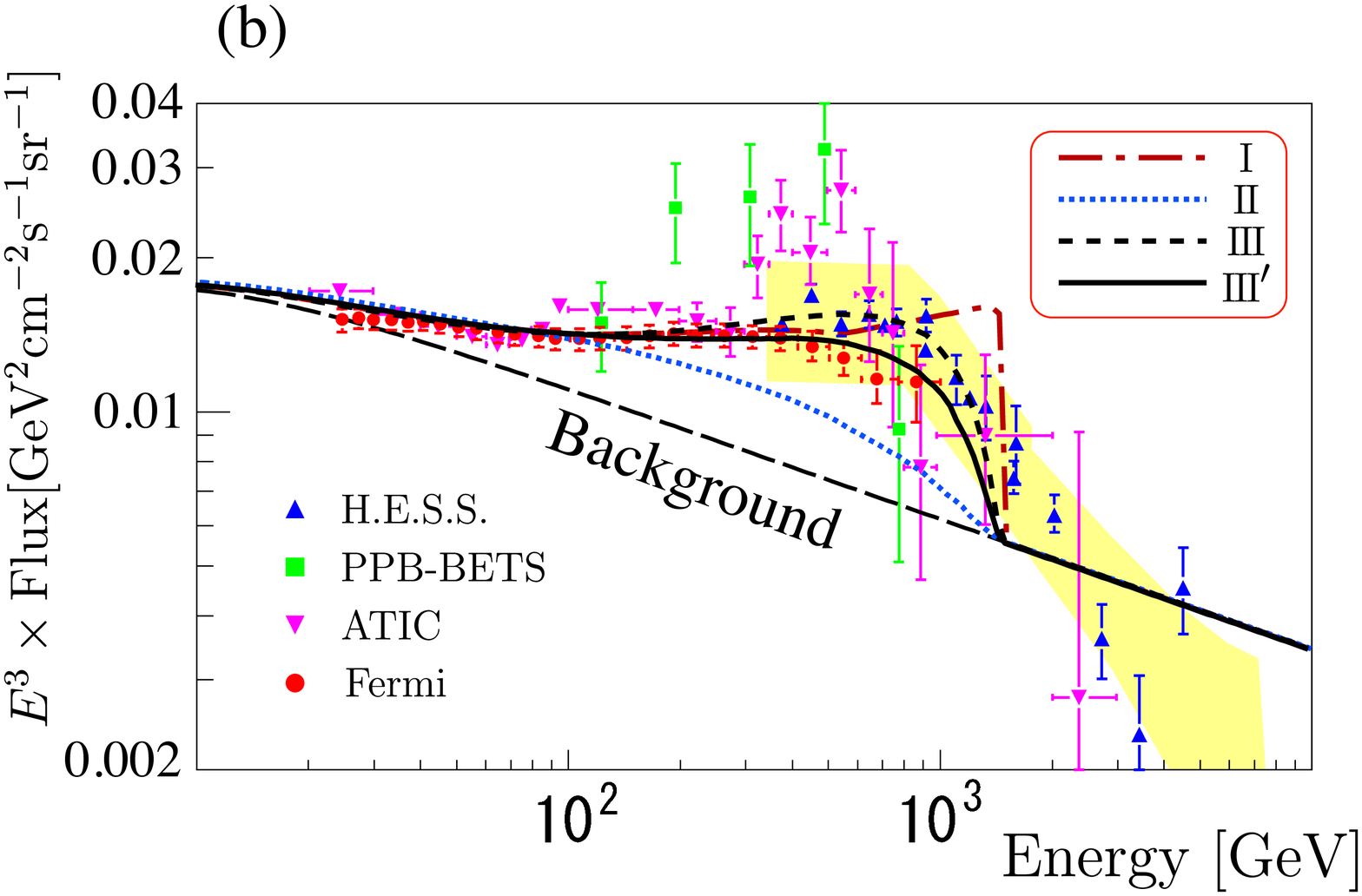 ,scale=.50,clip}\\
\end{center}
\caption{Cosmic ray signals in the present model.
(a): positron fraction with experimental data \cite{Adriani:2008zr,Aguilar:2007yf,Barwick:1997ig}.
(b): $(e^-+e^+)$ flux with experimental data~\cite{Collaboration:2009zk,Collaboration:2008aaa,ATIC,Torii:2008xu}.
The yellow zone shows a systematic error and the dashed line shows the background flux.
III$'$ is the same as the Case III except for DM's lifetime and branching fraction.
In this case, we set $\tau_{\rm DM}=9\times 10^{25}$ sec and the branching fraction 
of decay into $e$, $\mu$ and $\tau$ as 1:1:3, taking the mass ratio of stau to smuon (selectron)
as $\simeq (1/3)^{1/4}$.
}
\label{fig:signal}
\end{figure}

\begin{figure}[ht]
\begin{center}
\epsfig{file=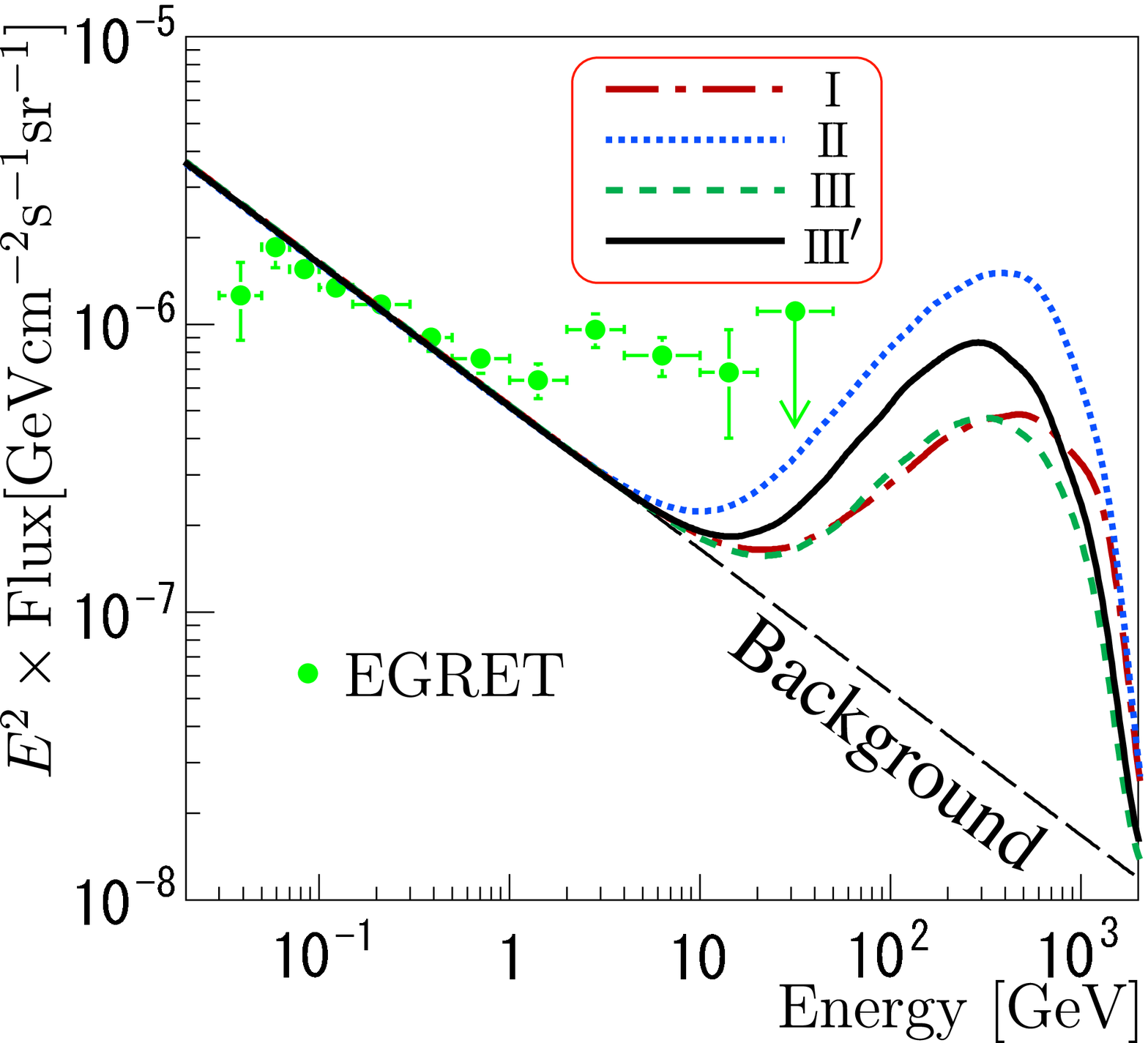 ,scale=.60,clip}\\
\end{center}
\caption{Predicted signals of diffuse gamma ray shown together 
with the EGRET data~\cite{Sreekumar:1997un,Strong:2004ry}.
}
\label{fig:signal_gamma}
\end{figure}

The electron and positron energy spectrum is estimated with the
program PYTHIA~\cite{Sjostrand:2006za}.  For the propagation of the
cosmic ray in the Galaxy, we adopt the same set-up in
Ref.~\cite{Shirai:2009kh} based on
Refs.~\cite{Hisano:2005ec,Ibarra:2008qg}, namely the MED diffusion
model~\cite{Delahaye:2007fr} and the NFW dark matter
profile~\cite{Navarro:1996gj}. As for the electron and positron
background, we have used the estimation given in
Refs.~\cite{Moskalenko:1997gh,Baltz:1998xv}, with a normalization
factor $k_{\rm bg} = 0.68$.  In Figs.~\ref{fig:signal} and
\ref{fig:signal_gamma}, we show the positron fraction, the electron
plus positron total flux and the diffuse gamma ray flux.

We can see from Fig.~\ref{fig:signal}, the above three cases nicely
explain the PAMELA result, while the case III seems to give a slightly
better fit to the Fermi and H.E.S.S. data with respect to the other
two cases. Note however that the fit to the data has an ambiguity due
to the relatively large uncertainties in the background estimation, as
well as possible astrophysical
contributions~\cite{Hooper:2008kg,Ioka:2008cv,Simet:2009ne,Grasso:2009ma,Shaviv:2009bu}.
As to the diffuse gamma-ray flux shown in Fig.~\ref{fig:signal_gamma},
the $\tau$ decay in the case II tends to give more contribution.  In
all the three cases we might be able to see some signatures from the
$\lambda_H$ decay in the diffuse gamma-rays in the future observation
with the Fermi satellite.

\section{Discussion and conclusions}
\label{sec:4}

So far we have assumed that the hidden gaugino $\lambda_H$ is the main
DM component.  However, as long as the R parity is conserved, the
neutralino LSP in the SSM also contributes to the DM density.  To
avoid the overproduction of the neutralino LSP, we assume either
neutralino-stau coannihilation or the Wino-like LSP.  In the former
case the stau mass must be close to the neutralino mass of ${\cal
  O}(100)$\,GeV, and the cosmic-ray spectra for such a mass spectrum
were studied in Ref.~\cite{Shirai:2009kh}. In the latter case, the
thermal relic abundance of the Wino LSP of mass ${\cal O}(100)$\,GeV
is smaller than the observed DM density~\footnote{ The Wino-like
  neutralino LSP can be realized in anomaly mediation~\cite{AMSB},
  which is feasible with more than two extra dimensions.  In this
  scheme, hidden matter multiplets charged under the hidden U(1) gauge
  symmetry must be introduced so that the hidden gaugino acquires a
  SUSY breaking mass.}, and we easily avoid the overproduction of the
LSP.

Let us discuss the production mechanism of the hidden gaugino,
$\lambda_H$, in the early universe.  Since the $\lambda_H$ has only
extremely suppressed interactions with the SSM particles, very high
reheating temperatures would be needed to generate a right amount of
$\lambda_H$ from thermal particle scatterings.\footnote{ Another
  possibility is non-thermal production from the inflaton
  decay. However, as shown in Ref.~\cite{Endo:2007ih}, an equal or
  even greater amount of the gravitino will be also generated in a
  similar process, and the situation will not be improved much. } This will be in
conflict with the big bang nucleosynthesis (BBN) constraint on the
gravitino abundance~\cite{Kawasaki:2008qe}.
 As
pointed out in Ref.~\cite{Shirai:2009kh}, one possible way to produce
$\lambda_H$ is to make use of the gravitino decay. In fact, the
gravitino must be heavier than $\lambda_H$, since otherwise the
$\lambda_H$ would promptly decay into the massless hidden gauge boson
and the gravitino. Therefore, the gravitino produced at the reheating
will decay into the hidden gaugino and gauge boson as well as the SSM
particles.

Let us estimate the $\lambda_H$ abundance from the gravitino decay.
To be concrete we consider two cases: $m_{3/2} = 10$\,TeV and
$100$\,TeV, since the BBN constraint on the gravitino abundance
depends on the gravitino mass.  In the former case, the gravitino
abundance can be as large as $Y_{3/2} \sim 10^{-13}$ without spoiling
the BBN result~\cite{Kawasaki:2008qe}.  The corresponding reheating
temperature is about $10^9$\,GeV, assuming the thermal gravitino
production.  The expected branching ratio of producing $\lambda_H$ is
${\cal O}(1)$\%, and the $\lambda_H$ abundance will be $\Omega_\lambda
h^2 \sim 10^{-3}$.  Note that the abundance of the neutralino LSP
produced through the gravitino decay does not have enough abundance to
explain the total DM density, and therefore the dominant contribution
must come from the thermal relic neutralino. This may be realized in
the neutralino-stau coannihilation region.
On the other hand, in the case of $m_{3/2} = 100$\,TeV, the gravitino
abundance can be as large as $Y_{3/2} \sim 10^{-12}$ for a reheating
temperature $T_R \sim 10^{10}$\,GeV.  The resultant $\lambda_H$
abundance will be $\Omega_\lambda h^2 \sim 10^{-2}$.  Interestingly,
the neutralino abundance from the gravitino decay is just a right
amount to explain the observed DM density $\Omega_{\lambda} h^2 \simeq
0.1$.\footnote{ The non-thermally produced neutralino will not
  annihilate efficiently even in the case of the Wino LSP, unless the
  gravitino mass is extremely large.  }  Thus, even if the $\lambda_H$
may not be the dominant component of DM, its fraction can be naturally
in the range of $1 - 10$\% depending on the gravitino mass and the
reheating temperature. The predictions on the cosmic-ray spectra
remain unchanged if we make the lifetime shorter correspondingly by
adopting a slightly smaller value of the B$-$L breaking scale $v_{\rm B-L}$.

In this paper we have studied representative decay processes in a
scenario that a hidden U(1) gaugino DM decays mainly through a mixing
with a U(1)$_{\rm B-L}$, producing energetic leptons. We have shown
that those energetic leptons from the DM decay can account for the
PAMELA and Fermi excesses in the cosmic-ray electrons/positrons.  The
predicted excess in the diffuse gamma-ray flux around several hundred
GeV can be tested by the Fermi satellite, and will provide us with
information on the decay processes. One of the merits of the current
scenario is that the gaugino in the SSM can be within the reach of
LHC, since at least one of the SSM neutralino lighter than the hidden
gaugino DM is necessary for the DM to decay.

\section*{Acknowledgement}
This work was supported by World Premier International Center
Initiative (WPI Program), MEXT, Japan.  The work of SS is supported in
part by JSPS Research Fellowships for Young Scientists.


\begin{thebibliography}{99}


\bibitem{Adriani:2008zr}
  O.~Adriani {\it et al.}  [PAMELA Collaboration],
  Nature {\bf 458}, 607 (2009)
  [arXiv:0810.4995 [astro-ph]].

\bibitem{Collaboration:2009zk}
  Fermi/LAT Collaboration,
  arXiv:0905.0025 [astro-ph.HE].




\bibitem{Collaboration:2008aaa}
  F.~Aharonian {\it et al.}  [H.E.S.S. Collaboration],
  Phys.\ Rev.\ Lett.\  {\bf 101}, 261104 (2008)
  [arXiv:0811.3894 [astro-ph]];
  arXiv:0905.0105 [astro-ph.HE].


\bibitem{Shirai:2009fq}
  S.~Shirai, F.~Takahashi and T.~T.~Yanagida,
  arXiv:0905.0388 [hep-ph].

\bibitem{Shirai:2009kh}
  S.~Shirai, F.~Takahashi and T.~T.~Yanagida,
  arXiv:0902.4770 [hep-ph].

\bibitem{Ibarra:2008kn}
  A.~Ibarra, A.~Ringwald, D.~Tran and C.~Weniger,
  arXiv:0903.3625 [hep-ph]; see also
 A.~Ibarra, A.~Ringwald and C.~Weniger,
  JCAP {\bf 0901}, 003 (2009)
  [arXiv:0809.3196 [hep-ph]].



\bibitem{Chen:2008yi}
  C.~R.~Chen, F.~Takahashi and T.~T.~Yanagida,
  Phys.\ Lett.\  B {\bf 671}, 71 (2009)
  [arXiv:0809.0792 [hep-ph]];
  Phys.\ Lett.\  B {\bf 673}, 255 (2009)
  [arXiv:0811.0477 [hep-ph]].
  C.~R.~Chen, M.~M.~Nojiri, F.~Takahashi and T.~T.~Yanagida,
  arXiv:0811.3357 [astro-ph].
  
  
\bibitem{Adriani:2008zq}
  O.~Adriani {\it et al.},
  Phys.\ Rev.\ Lett.\  {\bf 102}, 051101 (2009)
  [arXiv:0810.4994 [astro-ph]].






  




\bibitem{seesaw}
T.~Yangida, in Proceedings of the {\it{``Workshop on the Unified Theory and
 the Baryon Number in the Universe''}}, Tsukuba, Japan, Feb. 13-14, 1979, edited by
O.~Sawada and A.~Sugamoto, KEK report KEK-79-18, p. 95, 
and {\it{``Horizontal Symmetry And Masses Of Neutrinos''
}}, Prog. Theor. Phys. {\bf{64}} (1980) 1103;
M.~Gell-Mann, P.~Ramond and R.~Slansky, in {\it{``Supergravity''}}
 (North-Holland, Amsterdam, 1979) {\it{eds}}. D.~Z.~Freedom and P.~van 
Nieuwenhuizen, Print-80-0576 (CERN);
see also   P.~Minkowski,  Phys.\ Lett.\  B {\bf 67}, 421 (1977).


\bibitem{Arvanitaki:2008hq}
  A.~Arvanitaki, S.~Dimopoulos, S.~Dubovsky, P.~W.~Graham, R.~Harnik and S.~Rajendran,
  arXiv:0812.2075 [hep-ph].



\bibitem{Sjostrand:2006za}
  T.~Sjostrand, S.~Mrenna and P.~Skands,
  JHEP {\bf 0605}, 026 (2006)
  [arXiv:hep-ph/0603175].



\bibitem{Hisano:2005ec}
  J.~Hisano, S.~Matsumoto, O.~Saito and M.~Senami,
  Phys.\ Rev.\  D {\bf 73}, 055004 (2006)
  [arXiv:hep-ph/0511118].

   
  
\bibitem{Ibarra:2008qg}
  A.~Ibarra and D.~Tran,
  JCAP {\bf 0807}, 002 (2008)
  [arXiv:0804.4596 [astro-ph]].
  
  



\bibitem{Delahaye:2007fr}
  T.~Delahaye, R.~Lineros, F.~Donato, N.~Fornengo and P.~Salati,
  Phys.\ Rev.\  D {\bf 77}, 063527 (2008)
  [arXiv:0712.2312 [astro-ph]].
 
     
\bibitem{Navarro:1996gj}
  J.~F.~Navarro, C.~S.~Frenk and S.~D.~M.~White,
  Astrophys.\ J.\  {\bf 490}, 493 (1997)
  [arXiv:astro-ph/9611107].
  
  
\bibitem{Moskalenko:1997gh}
  I.~V.~Moskalenko and A.~W.~Strong,
  Astrophys.\ J.\  {\bf 493} (1998) 694
  [arXiv:astro-ph/9710124].


\bibitem{Baltz:1998xv}
  E.~A.~Baltz and J.~Edsjo,
  Phys.\ Rev.\  D {\bf 59} (1999) 023511
  [arXiv:astro-ph/9808243].

  

\bibitem{Aguilar:2007yf}
  M.~Aguilar {\it et al.}  [AMS-01 Collaboration],
  Phys.\ Lett.\  B {\bf 646}, 145 (2007)
  [arXiv:astro-ph/0703154].

\bibitem{Barwick:1997ig}
  S.~W.~Barwick {\it et al.}  [HEAT Collaboration],
  Astrophys.\ J.\  {\bf 482}, L191 (1997)
  [arXiv:astro-ph/9703192].




\bibitem{ATIC}
J.~Chang {\it et al.} [ATIC Collaboration],
Nature {\bf 456}, 362 (2008).

\bibitem{Torii:2008xu}
  S.~Torii {\it et al.},
  arXiv:0809.0760 [astro-ph].


\bibitem{Sreekumar:1997un}
  P.~Sreekumar {\it et al.}  [EGRET Collaboration],
  Astrophys.\ J.\  {\bf 494}, 523 (1998)
  [arXiv:astro-ph/9709257].

\bibitem{Strong:2004ry}
  A.~W.~Strong, I.~V.~Moskalenko and O.~Reimer,
  Astrophys.\ J.\  {\bf 613}, 956 (2004)
  [arXiv:astro-ph/0405441].
    
\bibitem{Hooper:2008kg}
  D.~Hooper, P.~Blasi and P.~D.~Serpico,
  JCAP {\bf 0901}, 025 (2009)
  [arXiv:0810.1527 [astro-ph]];
  H.~Yuksel, M.~D.~Kistler and T.~Stanev,
  arXiv:0810.2784 [astro-ph];
  S.~Profumo,
  arXiv:0812.4457 [astro-ph].

  
\bibitem{Ioka:2008cv}
  K.~Ioka,
  arXiv:0812.4851 [astro-ph].
  

\bibitem{Simet:2009ne}
  M.~Simet and D.~Hooper,
  arXiv:0904.2398 [astro-ph.HE].


\bibitem{Grasso:2009ma}
  D.~Grasso {\it et al.}  [FERMI-LAT Collaboration],
  arXiv:0905.0636 [astro-ph.HE].



\bibitem{Shaviv:2009bu}
  N.~J.~Shaviv, E.~Nakar and T.~Piran,
  arXiv:0902.0376 [astro-ph.HE];
  P.~Blasi,
  arXiv:0903.2794 [astro-ph.HE];
  P.~L.~Biermann, J.~K.~Becker, A.~Meli, W.~Rhode, E.~S.~Seo and T.~Stanev,
  arXiv:0903.4048 [astro-ph.HE];
  Y.~Fujita, K.~Kohri, R.~Yamazaki and K.~Ioka,
  arXiv:0903.5298 [astro-ph.HE].

 

   \bibitem{AMSB} 
  L.~Randall and R.~Sundrum,
  Nucl.\ Phys.\ B {\bf 557}, 79 (1999);\\
  G.~F.~Giudice, M.~A.~Luty, H.~Murayama and R.~Rattazzi,
  JHEP {\bf 9812}, 027 (1998);\\
  J.~A.~Bagger, T.~Moroi and E.~Poppitz,
  JHEP {\bf 0004}, 009 (2000).




  
\bibitem{Endo:2007ih}
  M.~Endo, F.~Takahashi and T.~T.~Yanagida,
  Phys.\ Lett.\  B {\bf 658}, 236 (2008)
  [arXiv:hep-ph/0701042];
  Phys.\ Rev.\  D {\bf 76}, 083509 (2007)
  [arXiv:0706.0986 [hep-ph]].
  
  
  
\bibitem{Kawasaki:2008qe}
  M.~Kawasaki, K.~Kohri, T.~Moroi and A.~Yotsuyanagi,
  Phys.\ Rev.\  D {\bf 78}, 065011 (2008)
  [arXiv:0804.3745 [hep-ph]].



  

\end{thebibliography}
\end{document}